\documentclass[prl,superscriptaddress,superscriptaddress,%
twocolumn,showpacs,preprintnumbers,amsmath,amssymb]{revtex4-1}
\usepackage{graphicx}
\usepackage[breaklinks, colorlinks=true, pdfstartview=FitV,%
 linkcolor=red, citecolor=blue, urlcolor=blue]{hyperref}
\usepackage{color}

\newcommand{\calA}{\mathcal{A}}
\newcommand{\calB}{\mathcal{B}}
\newcommand{\calC}{\mathcal{C}}
\newcommand{\calM}{\mathcal{M}}
\newcommand{\calN}{\mathcal{N}}
\newcommand{\calX}{\mathcal{X}}
\newcommand{\bk}{\boldsymbol{k}}
\newcommand{\bj}{\boldsymbol{j}}
\newcommand{\bx}{\boldsymbol{x}}
\newcommand{\bB}{\boldsymbol{B}}

\newcommand{\Nc}{N_{\text{c}}}
\newcommand{\Nf}{N_{\text{f}}}
\newcommand{\bara}{\bar{a}}

\begin{document}

\title{Spatial modulation and topological current
       in holographic QCD matter}

\author{Kenji Fukushima}
\affiliation{Department of Physics, Keio University,
             Kanagawa 223-8522, Japan}

\author{Pablo A.\ Morales}
\affiliation{Department of Physics, Keio University,
             Kanagawa 223-8522, Japan}

\begin{abstract}
  We investigate an impact of the axial-vector interaction on spatial
  modulation of quark matter.  A magnetic field coupled with baryon
  density leads to a topological axial current, so that the effect of
  the axial-vector interaction is crucially enhanced then.  Using the
  Sakai-Sugimoto model we have found that, contrary to a na\"{i}ve
  expectation, the spatially modulated phase is less favored for a
  stronger magnetic field, which is realized by the presence of
  topological current.
\end{abstract}
\maketitle


\paragraph{Introduction}
The phase diagram of hot and dense matter out of quarks and gluons has
not been clarified satisfactorily based on the first-principle theory
of the strong interaction, i.e.\ quantum chromodynamics (QCD).  The
most severe obstacle is the notorious sign problem of the Dirac
determinant at finite quark density $\rho$ or chemical potential
$\mu$, which prevents us from the direct application of the
Monte-Carlo simulation in the region with $\mu\gtrsim
T$~\cite{Fukushima:2010bq}.

Instead of the lattice simulation, one could have deduced possible
phase structures using chiral effective models, see
e.g.\ \cite{Herbst:2013ail} for a recent work.  It is conjectured from
model studies that the chiral phase transition might be of first order
at high density, so that a second-order critical point called the QCD
critical point~\cite{Asakawa:1989bq} could appear on the phase
diagram, the discovery of which is one of the major goals of the
beam-energy scan program in heavy-ion collision
experiments~\cite{Kumar:2012fb}.  The model setup, however, suffers
from uncontrolled uncertainties and the QCD critical point is a
model-dependent prediction.  It is well understood by now that the
vector-type interaction $\sim (\bar{\psi}\gamma_\mu\psi)^2$, 
which gives rise to the density-density interaction $\sim \rho^2$ even
in the mean-field level, crucially affects the liquid-gas phase
transition of dense quark
matter~\cite{Kitazawa:2002bc,Fukushima:2012mz} (see also
\cite{Contrera:2012wj}).  Moreover, nowadays, spatially inhomogeneous
states are becoming a more and more realistic candidate that may
supersede the conventional first-order phase
boundary~\cite{Deryagin:1992rw}, which is rather robust against the
vector interaction~\cite{Fukushima:2012mz,Carignano:2010ac}.

The simplest Ansatz to introduce spatial modulation is the chiral
spiral or the dual chiral-density wave,
\begin{equation}
 \langle\bar{\psi}\psi\rangle = \Delta\cos(\bk\cdot\bx)\;,\quad
 \langle\bar{\psi}\gamma_5\tau^3\psi\rangle
  = \Delta\sin(\bk\cdot\bx)\;,
\end{equation}
which is reminiscent of the $p$-wave $\pi^0$ condensate in symmetric
nuclear matter.  Recalling the history of the pion
condensation~\cite{pion}, one may well consider that a
spin-isospin short-range interaction could significantly diminish the
reality of chiral spirals;  it was indeed the case for the pion
condensation that is disfavored by the so-called Landau-Migdal
parameters $g'$ associated with short-range effective interaction in
Fermi liquid theory (see also \cite{Tatsumi:2003fa} for some arguments
in favor of the pion condensation).  In the relativistic language,
thus, it is conceivable that the axial-vector interaction
$\sim (\bar{\psi}\gamma_5 \gamma_\mu \boldsymbol{\tau}\psi)^2$ may be 
influential on spatial modulation of quark matter, though the vector
interaction is not.  This is an important question but, to the best of
our knowledge, there is no theoretical investigation on this issue.
The difficulty lies in the fact that the axial-vector has no
mean-field contribution unlike the density in the vector channel, and
therefore one should go beyond the mean-field approximation.  So far,
the renormalization-group improvement has been successful for the
homogeneous states only~\cite{Herbst:2013ail}.

This situation would be drastically changed if we turn an external
magnetic field $B$ on.  Such a system of dense quark matter at strong
magnetic field has been intensely investigated.  It was pointed out
first in the Sakai-Sugimoto model~\cite{Sakai:2004cn} which is a
holographic dual of large-$\Nc$ QCD that $B$ lowers the critical
$\mu$~\cite{Preis:2010cq}.  This observation turns out to be generic
in chiral models~\cite{Preis:2012fh} and is often referred to as the
inverse magnetic catalysis in contrast to the enhancement of chiral
symmetry breaking at zero density~\cite{Gusynin:1994re}.  In this way,
clarification of the QCD phase diagram along larger-$B$ direction is
an intriguing subject and many studies have been devoted to
it~\cite{Gatto:2012sp}.

There are also some theoretical works focused on inhomogeneous states
of dense quark matter at finite $B$:  In the strong-$B$ limit quarks
are dimensionally reduced into a (1+1)-dimensional system, so that the
ground-state structure should be a chiral spiral, i.e.\ chiral
magnetic spiral~\cite{Basar:2010zd}.  It is also possible that another
spiral can develop due to the presence of $B$~\cite{Ferrer:2012zq}.
In view of such results, it should be a natural expectation that a
stronger $B$ may ease a barrier to form spirals.

Here, in this work, we would address one important physical effect
that has been overlooked in these preceding works.  That is, the
inevitable generation of the topological current,
\begin{equation}
 \bj_A = \Nc \sum_f \frac{q_f^2 \mu}{2\pi^2}\bB \;,
\label{eq:cse}
\end{equation}
having the origin in quantum anomaly~\cite{Metlitski:2005pr}, should
be incorporated.  $\Nc$
is the number of color, $f$ runs over flavor degrees of freedom, and
$q_f$ is the electric charge of flavor $f$.  Interestingly, if
$\bj_A\neq0$ at finite $\mu$ and $B$, the axial-vector interaction has
a mean-field contribution $\bj_A^2$ in the same way as $\rho^2$
emerging from the vector interaction, which could have played a role
similar to the Landau-Migdal interaction and thus disfavored spirals
contrary to the na\"{i}ve expectation.  Although there are countless
works to study such chiral magnetic and separation effects as in
Eq.~\eqref{eq:cse}, nobody has ever considered its impact on the phase
structure at finite $\mu$ and $B$.

For the purpose to address these issues, the Sakai-Sugimoto model
suits the best.  We could use conventional methods, but
then it is difficult to quantify the axial-vector interaction.  There
is no such ambiguity in the holographic approach.  Besides, the
holographic technique for the phase diagram research has been
successfully advanced recently and the instability toward spatially
modulated phase has been discovered~\cite{Ooguri:2010xs}.  In the
presence of chiral chemical potential, also, similar instability
leading to a spiral has been identified in the Sakai-Sugimoto
model~\cite{BallonBayona:2012wx}.
\vspace{0.5em}


\paragraph{Holographic Description}
The gauge/gravity (or generally bulk/boundary) correspondence states
that the full quantum generating functional of 4-dimensional field
theory is equivalent to the on-shell action of the gravity theory with
corresponding source at the ultraviolet (UV) boundary.  Thus, $\Nc$ D4
branes compactified along the $x_4$-direction represent the gluonic
degrees of freedom~\cite{Witten:1998zw} and $\Nf$
${\rm D8}$-$\overline{\rm D8}$ branes realize the spontaneous breaking
of ${\rm U}(\Nf)_{\rm L}\times{\rm U}(\Nf)_{\rm R}$ chiral symmetry in
QCD~\cite{Sakai:2004cn}.  In the same way as in the first paper of
\cite{Ooguri:2010xs} we focus on the situation where ${\rm D8}$ and
$\overline{\rm D8}$ are separate above the deconfinement transition.
There, the induced-metric on the flavor branes is,
\begin{equation}
 ds^2 = u^{3/2}\bigl[ f(u) d\tau^2 + d\bx^2 \bigl]
  + \Bigl[ u^{3/2} x_4'(u)^2 + \frac{1}{u^{3/2} f(u)} \Bigr] du^2 \;,
\label{eq:metric}
\end{equation}
where $f(u)=1-u_T^3/u^3$.  We note that all variables are made
dimensionless by the AdS radius.  The horizon at $u=u_T$ defines
the Hawking temperature, which is translated to the physical
temperature as $T = 3u_T^{1/2}/(4\pi)$.  In the chiral symmetric phase
D8 and $\overline{\rm D8}$ are simply straight, so that ${x_4'(u)}=0$
is chosen.

Then, the DBI action in the flavor sector can be expressed with the
metric from Eq.~\eqref{eq:metric} and the ${\rm U}(1)$ field strength
tensor $F_{\alpha\beta}$ which is split into $B$ in the $z$-direction
(under simplification that all $\Nf$ flavors have the same electric
charge), the background $\bara_0$ and $\bara_z$ corresponding to $\mu$
and $j_A^z$, and spatially inhomogeneous fluctuations
$f_{\alpha\beta}$.  The 5-dimensional effective action reads,
\begin{equation}
 \begin{split}
 S_{\text{D8}}^{\text{DBI}} &= \mathcal{N}\int d\tau\,d^3 x\,du\,
  u^{1/4}\sqrt{-\det(g_{\alpha\beta}+F_{\alpha\beta})} \\
  &= \calN\int d\tau\, d^3 x\, du\, u^{5/2}
  \sqrt{\calA\cdot\calB}\;(1+\calX)
 \end{split}
\end{equation}
with an overall (irrelevant) constant $\calN$ and
\begin{equation}
 \calA = 1 - \bara_0'(u)^2 + f(u) \bara_z'(u)^2 \;, \quad
 \calB = 1 + B^2 u^{-3} \;,
\label{eq:A_B}
\end{equation}
and the fluctuation part $\calX$ up to the quadratic order with
respect to $f_{xy}=\partial_x a_y-\partial_y a_x$, and $f_{yz}$,
$f_{zx}$, $f_{ux}$, $f_{uy}$, $f_{uz}$ with similar definitions.

Hence, together with the Chern-Simons action,
$S^{\text{CS}}=(\mathcal{N}/8)\int d\tau\,d^3 x\,du\,
\epsilon^{\mu_1\mu_2\mu_3\mu_4\mu_5}A_{\mu_1}F_{\mu_2\mu_3}F_{\mu_4\mu_5}$,
we can define variables conjugate to $\bara_0'$ and $\bara_z'$
using the full action $S=S_{\text{D8}}^{\text{DBI}}+S^{\text{CS}}$, as
\begin{align}
 \rho &= -\frac{\delta S}{\delta \bara_0'(u)} = u^{5/2} \bara_0'(u)
 \sqrt{\frac{\calB}{\calA}} - 3 B \bara_z(u) \;,
\label{eq:rho}\\
 b &= \frac{\delta S}{\delta \bara_z'(u)} = u^{5/2} f(u) \bara_z'(u)
  \sqrt{\frac{\calB}{\calA}} - 3 B \bara_0(u) \;.
\label{eq:b}
\end{align}
Because $S$ is dependent on not $\bara_0$ and $\bara_z$ but $\bara_0'$
and $\bara_z'$ only, $\rho$ and $b$ fixed from the equations of motion
are $u$-independent.  We find $b=0$ by evaluating it at $u=u_T$, and
from the boundary condition $\bara_0(\infty)=\mu$, we can get the
asymptotic forms as
\begin{equation}
 \bara_z(u) \simeq -2\mu B u^{-3/2} \;,\quad
 \bara_0(u) \simeq \mu-\frac{9}{8}\rho u^{-3/2} \;,
\end{equation}
near the UV boundary ($u\sim\infty$).  This asymptotic behavior of
$\bara_z(u)$ represents the topological vector and axial-vector
currents~\eqref{eq:cse}~\cite{Preis:2010cq,Preis:2012fh,%
BallonBayona:2012wx,Yee:2009vw}.  In our numerical calculations we
fully solve Eqs.~\eqref{eq:rho} and \eqref{eq:b} for a given density
$\rho$ to obtain the whole profile of $\bara_0(u)$ and $\bara_z(u)$.

From the concrete form of $\calX$ we can get the equations of motion
with respect to fluctuations $a_i$ ($i=x,y,z$) as
\begin{align}
 & u^{-1/2}\!\sqrt{\frac{\calA}{\calB}} \Bigl(
  \frac{\partial_y f_{yx}}{\calB} \!+\! \calC \partial_z f_{zx} \Bigr)
  + \partial_u\! \Bigl[ \frac{u^{5/2} f(u) f_{ux}}{\sqrt{\calA
  \cdot \calB}} \Bigr] \notag\\
 &\qquad\qquad\qquad\qquad\qquad\qquad\qquad\quad\:\:\:
  + 3 \bara_0' f_{yz} = 0 \;,
\label{eq:eom1}\\
 & u^{-1/2}\!\sqrt{\frac{\calA}{\calB}} \Bigl(
  \frac{\partial_x f_{xy}}{\calB} \!+\! \calC \partial_z f_{zy} \Bigr)
  + \partial_u\! \Bigl[ \frac{u^{5/2} f(u) f_{uy}}{\sqrt{\calA
  \cdot \calB}} \Bigr] \notag\\
 &\qquad\qquad\qquad\qquad\qquad\qquad\qquad\quad\:
  + 3 \bara_0' f_{zx} = 0 \;,
\label{eq:eom2}\\
 & u^{-1/2}\!\sqrt{\frac{\calA}{\calB}}\, \calC \bigl(
  \partial_x f_{xz} + \partial_y f_{yz} \bigr) + \partial_u
  \Bigl[ u^{5/2}\!\sqrt{\frac{\calB}{\calA}}\,\calC f(u) f_{uz}
  \Bigr] \notag\\
 &\qquad\qquad\qquad\qquad\qquad\qquad\qquad\quad\:
  + 3 \bara_0' f_{xy} = 0 \;,
\label{eq:eom3}
\end{align}
where $\calC = 1 - f(u) \bara_z'(u)^2 / \calA$.
\vspace{0.5em}

\begin{figure}
 \includegraphics[width=0.95\columnwidth]{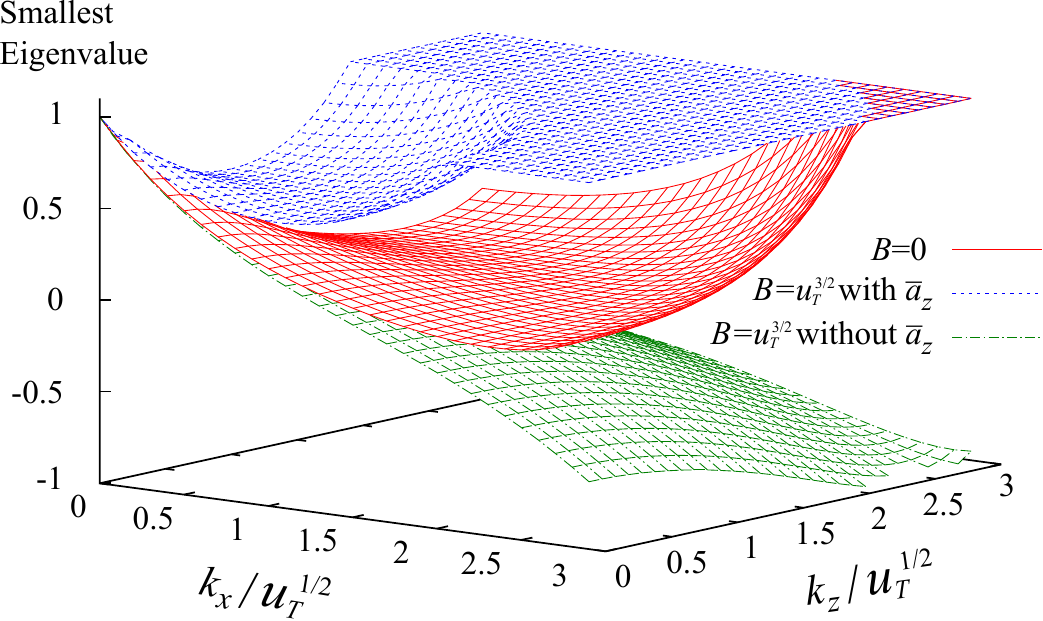}
 \caption{Smallest eigenvalue of the matrix $\calM$ as a function of
   $k_x$ (perpendicular to $B$) and $k_z$ (parallel to $B$) at
   $\rho=3.72u_T^{5/2}$ at $B=0$ (surface in the middle) and at
   $B=u_T^{3/2}$ with $\bara_z$ (surface in the top) and without
   $\bara_z$ (surface in the bottom).}
 \label{fig:critical}
\end{figure}


\paragraph{Numerical Results}
A finite $B$ breaks rotational symmetry and we cannot find the
eigenmodes as done in \cite{Ooguri:2010xs}.  Let us here explain how
to proceed to numerical analyses.  Our goal is to locate the critical
$\rho$ or $\mu$ (denoted by $\mu_c$ hereafter) at which
Eqs.~\eqref{eq:eom1}, \eqref{eq:eom2}, and \eqref{eq:eom3} have
normalizable solutions with some momenta $k_x$, $k_y$, $k_z$ in
Fourier space.  In fact the normalizability condition or the boundary
conditions $a_i(\infty)\to0$ dictate how the energy dispersion
relations behave.  Since we drop time dependence, our solutions
describe the dispersion relation at zero energy.  If a zero-energy
excitation is realized with non-zero momenta, a homogeneous state
should become unstable.

To solve three differential equations for $a_i$ from $u=u_T$ to
$u=\infty$, we need to specify the initial condition for $a_i'(u_T)$.
These are uniquely taken if we require the solutions to be
non-singular at $u=u_T$;  since $f(u)$ vanishes at $u=u_T$, only the
term with $\partial_u$ acting on $f(u)$ remains non-zero unless
$a_i''(u_T)$ is singular.  Then, we can easily express $a_i'(u_T)$
using $a_i(u_T)$.  For example, we can deduce $a_x'(u_T)$ from
Eq.~\eqref{eq:eom1} as
\begin{align}
 a_x'(u_T) &= \frac{\calA}{3u_T^2}\Bigl[\frac{k_y^2 a_x
  -k_x k_y a_y}{\calB} + \calC \bigl( k_z^2 a_x - k_z k_x a_z\bigr)
  \Bigr] \notag\\
 &\qquad\qquad
  -i \sqrt{\calA\cdot\calB}\, \bara_0' u_T^{-3/2}(k_y a_z - k_z a_y) \;,
\end{align}
as well as $a_y'(u_T)$ and $a_z'(u_T)$ similarly.

Now, we are ready for solving Eqs.~\eqref{eq:eom1}, \eqref{eq:eom2},
and \eqref{eq:eom3} numerically, and the final values $a_i(\infty)$
are then given as functions of the initial values $a_i(u_T)$, which
can be expressed, thanks to the linearity, as follows;
\begin{equation}
 \begin{pmatrix} a_x(\infty) \\ a_y(\infty) \\ a_z(\infty)
 \end{pmatrix}
 = \calM \begin{pmatrix} a_x(u_T) \\ a_y(u_T) \\ a_z(u_T)
  \end{pmatrix} \;,
\end{equation}
where $\calM$ is a $3\times3$ matrix, having three eigenvalues.  If an
eigenvalue turns out to be vanishing at some momenta, the initial
condition set with the corresponding eigenvector leads to the desired
boundary conditions, $a_x(\infty)=a_y(\infty)=a_z(\infty)=0$.

Figure~\ref{fig:critical} shows the smallest eigenvalue of $\calM$ as
a function of $k_x$ and $k_z$ (we can set $k_y=0$ without loss of
generality).  We can get rid of $u_T$-dependence by rescaling $\rho$,
$\mu$, $B$, and $k_i$.  We find that $\rho=3.72u_T^{5/2}$ is the
critical value for $B=0$ at which the smallest eigenvalue touches zero
at $|\boldsymbol{k}|=2.3u_T^{1/2}$ (which confirms
\cite{Ooguri:2010xs}).  When we increase $B$, the smallest eigenvalue
is pushed up as depicted by the upper surface in
Fig.~\ref{fig:critical}, and thus the critical density should get
larger.  This means that a larger $B$ disfavors the spatially
modulated phase.  Though it is not visually clear from
Fig.~\ref{fig:critical}, the eigenvalue is slightly tilted in the
presence of $B$ and the minimum of the eigenvalues is located on
$k_x\neq0$ and $k_z=0$.

\begin{figure}
 \includegraphics[width=0.85\columnwidth]{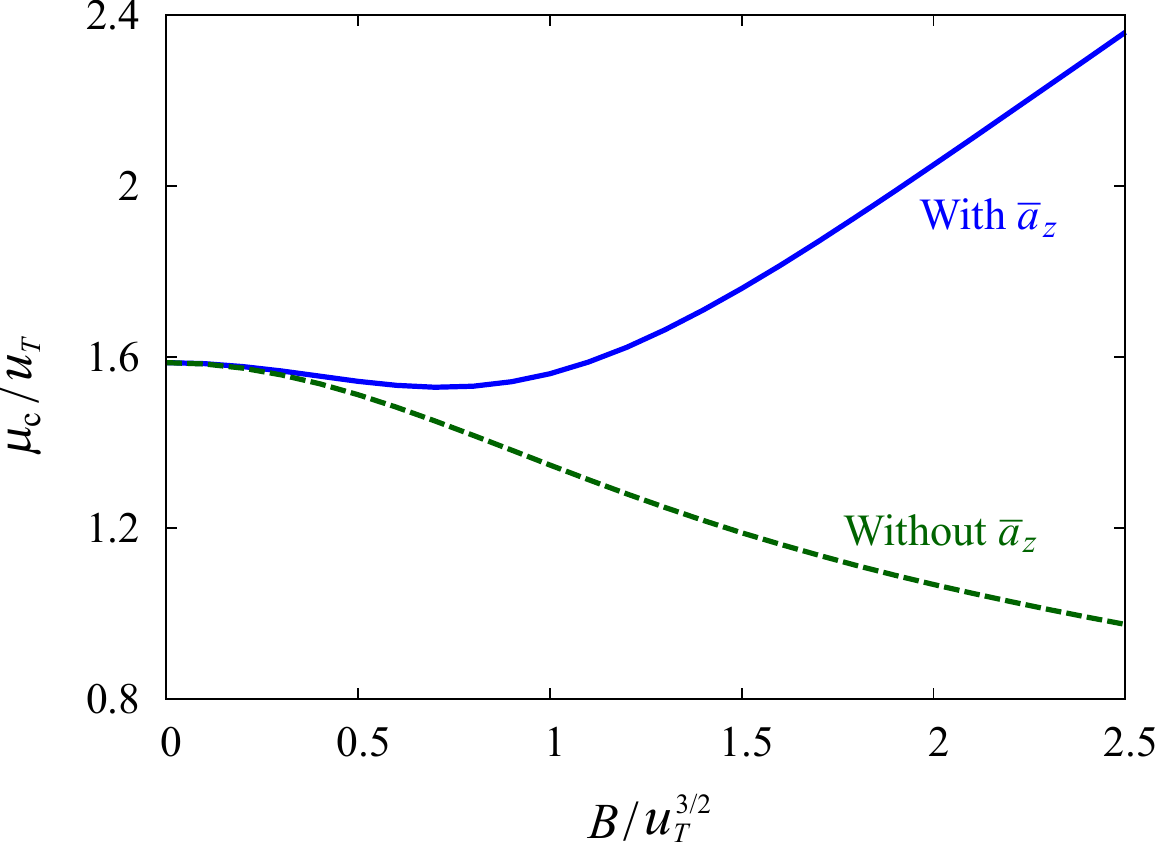}
 \caption{Critical chemical potential $\mu_c$ as a function of $B$.
   The upper solid curve represents the result with $\bara_z$ taken
   into account and the lower dashed line represents the result
   without $\bara_z$.}
 \label{fig:changeB}
\end{figure}

In terms of the chemical potential the relation between $\mu_c$ and
$B$ is more complicated.  As seen by the solid curve in
Fig.~\ref{fig:changeB} $\mu_c$ rather goes down with increasing $B$ as
long as the magnetic field is small enough, $B/u_T^{3/2}\lesssim 1$,
even though the critical $\rho$ monotonically grows up.  This is
simply because the phase space is enhanced by $B$;  if $B$ is raised
up for a fixed $\mu$, the corresponding density $\rho$ becomes
larger.
\vspace{0.5em}

\paragraph{Discussions}
It could have been more intuitively understandable if $B$ favored more
modulation in view of the chiral magnetic spirals at $B\to\infty$.
Here, in order to think of the effect of the topological
current~\eqref{eq:cse}, let us drop $\bara_z(u)$ off from the
calculation.  Of course, $\bara_z(u)=0$ is \textit{not} a solution of
the equation of motion, but this artificial manipulation in the
present holographic treatment can mimic the common approximation to
neglect $\bj_A$ in most non-holographic calculations.

In this case without $\bara_z$ we find that the smallest eigenvalue is
significantly pushed down by $B$ as seen in the bottom surface in
Fig.~\ref{fig:critical}.  This indicates that the critical density is
lowered by $B$ which makes a sharp contrast to the case with
$\bara_z$ (and thus $\bj_A$).  Needless to say, the critical chemical
potential $\mu_c$ also exhibits an opposite behavior to the previous
case with $\bara_z$, which is evident from the dashed curve in
Fig.~\ref{fig:changeB}.

In the holographic approach, generally, it is hard to carve distinct
physical effects out from the final results, and we did not spell out
the axial-vector interaction
$\sim (\bar{\psi}\gamma_5 \gamma_\mu \boldsymbol{\tau}\psi)^2$.
Nevertheless, our finding based on the comparison with and without
$\bara_z$ is suggestive enough to demonstrate the importance of the
axial-vector interaction along the same direction as the Landau-Migdal
interaction disfavoring the $p$-wave pion condensation.  It is an
intriguing future problem to implement the axial-vector interaction
in conventional chiral models such as the (Polyakov-loop coupled)
Nambu--Jona-Lasinio model and the quark-meson model to confirm our
finding and elucidate more microscopic dynamics.  In fact, in these
chiral models, $\bj_A$ should be treated as a mean-field variable and
$\bj_A$ is ``renormalized'' then~\cite{Fukushima:2010zza}.
Similar corrections on the topological current are reported also with
explicit QED calculations~\cite{Gorbar:2013upa}.
\vspace{0.5em}

\begin{figure}
 \includegraphics[width=0.85\columnwidth]{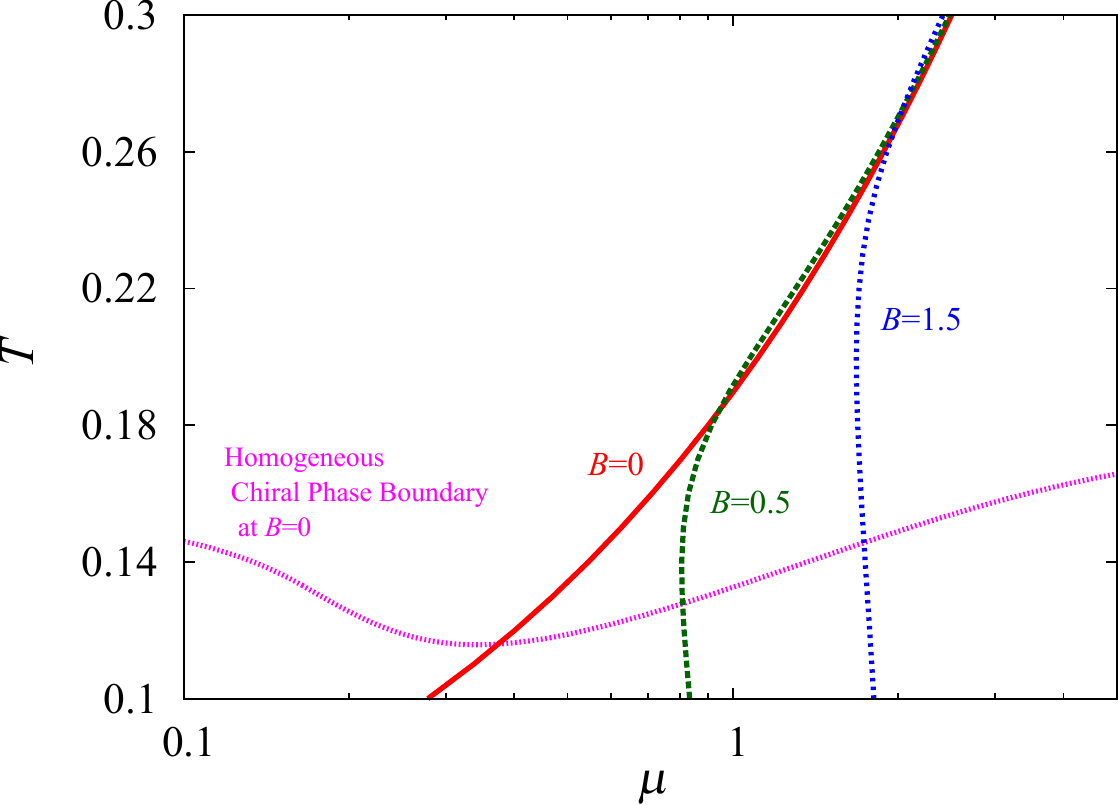}
 \caption{Phase boundaries of the onset of the spatially modulated
   phase at $B=0$ (solid curve), $B=0.5$ (dashed curve), and $B=1.5$
   (dotted curve) in the unit of not $u_T$ but the AdS radius.  For
   reference the phase boundary for the homogeneous chiral
   transition~\cite{Bergman:2007wp} is also shown.}
 \label{fig:boundary}
\end{figure}


\paragraph{Summary}
We calculated the critical density and the critical chemical potential
$\mu_c$ for spatial modulation at finite $B$.  We found that the
spatial modulation is disfavored for a larger $B$, which becomes
manifest on the phase diagram as summarized in
Fig.~\ref{fig:boundary}.  When $B=0$, we can find
$\mu_c\simeq 1.59u_T=27.9T^2$ that draws a solid curve in
Fig.~\ref{fig:boundary} (as seen in \cite{Ooguri:2010xs}).  This phase
boundary is shifted toward larger $\mu$ with increasing $B$, so that a
stronger $B$ causes shrinkage of the region with spatial inhomogeneity
on the phase diagram.  The effect of $B$ appears tamed at higher $T$,
which can be explained from Eq.~\eqref{eq:A_B} in which $B^2/u_T^3$
becomes negligible for high $T$ and thus large $u_T$.  By comparing
the results with and without the background $\bara_z(u)$, we conclude
that the disfavor of spatially modulated phase at finite $B$ is
attributed to the topological currents and presumably the axial-vector
interaction strengthened by $\bj_A$.

We are now making progress to explore the whole structure of the
holographic QCD phase diagram at finite $T$, $\mu$, and $B$ including
the effect of spontaneous chiral-symmetry breaking and baryon density
source that both make $x_4(u)$ take a non-trivial shape.  This will be
reported elsewhere.
\vspace{0.5em}

\acknowledgments
  We thank S.~Nakamura, I.~Shovkovy, and W.~Weise for useful
  discussions.
  K.~F.\ was supported by JSPS KAKENHI Grant \# 24740169.

\end{document}